\documentclass{article}
\usepackage{fullpage}
\usepackage{amsfonts,amssymb}
\usepackage{amsthm}
\usepackage{tikz}
\usetikzlibrary{calc}
\usetikzlibrary{arrows}
\usepackage{tikz-3dplot}
\usepackage{color} 
\usepackage[fleqn]{amsmath}
\usepackage{amsthm}
\usepackage{amssymb}
\usepackage{amsfonts}
\usepackage{bbm}
\usepackage{epsfig}
\usepackage{times}
\usepackage{hyperref}
\usepackage{bm}
\usepackage{float} 
\usepackage{appendix} 
\usepackage{lscape}
\usepackage{cite}
\usepackage{mathrsfs}
\usepackage{appendix}
\usepackage{setspace}
\usepackage{mathtools}
\usepackage{mdframed}

\theoremstyle{definition}

\usepackage{authblk}

\begin{document}
	
	\title{Operational Theories as Structural Realism}
	
	\author[1]{Emily Adlam} 
	\date{\today} 
	\affil[1]{The University of Western Ontario}

	\maketitle
	
	\abstract{We undertake a reconstruction of the epistemic significance of research on operational theories in quantum foundations. We suggest that the space of operational theories is analogous to the space of possible worlds employed in the possible world semantics for modal logic, so research of this sort can be understood as probing modal structure. Thus we argue that operational axiomatisations of quantum mechanics may be interpreted as a novel form of structural realism; we discuss the consequences of this interpretation for the philosophy of structural realism and the future of operational theories. 
	}

	\section{Introduction} 
 
In  the field of quantum foundations there is a thriving research programme which involves placing quantum mechanics in a wider space of operationally defined theories in order to gain insight into its structure. This research generally involves proving theorems of the form `every theory which has the feature(s) $\{ A \}$ must also have the feature(s) $\{ B \}$,' where often $\{ B\}$ is taken to be characteristic of quantum mechanics while $\{ A\}$ is described as \emph{ `simple,'}\cite{Masanes_2011} \emph{`natural,'} \cite{Masanes_2011} \emph{ `elementary,'}\cite{CDK} or \emph{`reasonable,'}\cite{Hardyreasonable} and the result is advertised as providing \emph{`insight into the reasons why quantum theory is the way it is.'}\cite{Hardyreasonable} Indeed, it is sometimes suggested that we should cease attempting to interpret quantum mechanics and instead seek to axiomatize it in this way\cite{10.1086/525620}. For example, Rovelli tells us that \emph{`quantum mechanics will cease to look puzzling only when we will be able to derive the formalism of the theory from a set of simple physical assertions (“postulates”, “principles”) about the world.'}\cite{1996IJTP...35.1637R} 

While   the intuition that we learn something interesting from these sorts of axiomatisations seems reasonable, it is by no means obvious \emph{what} precisely we learn. To improve the situation, it is necessary to give an account of the epistemic and/or ontological status of the postulates in order to say what precisely an axiomatisation of quantum mechanics is supposed to achieve. There are various existing philosophical analyses of the axiomatisation research programme, but many of these analyses have a strongly instrumentalist or even idealist flavour - for example, it has been suggested that we should understand these results as a demonstration that physics is really about information\cite{Wheeler1989-WHEIPQ, D_Ariano_2016,Masanes_2013, Chiribella} or language\cite{Grinbaum_2017}.  Here we offer an interpretation which is more compatible with scientific realism, arguing that the operational theories approach can be regarded as a form of structural realism. It follows that the detailed framework that has been developed by researchers in the field may potentially be  useful to structural realists as a way of clarifying and elaborating their epistemic commitments. Thus we will argue that the operational theories framework has valuable philosophical applications which have not previously been appreciated.

In section \ref{OT} we give some background on operational theories and explain the main obstacles to a realist interpretation of this research. In section \ref{counterfactual} we discuss the difference between axiomatisation and interpretation; then in section \ref{structure} we set out an account of the counterfactual features of the operational theory framework as encoding statements about the modal structure of reality, and argue that the process of axiomatisation should be understood as a search for the axiomatisation which best expresses this unique modal structure. In section \label{cosq} we make some interpretational remarks, and  finally, in section \ref{future}, we make some comments about the outlook for this approach. 

\section{Operational Theories \label{OT}}

Quantum mechanics is unique amongst theories in that it is more a \emph{recipe} for theories than a theory itself: it is a general theoretical prescription which can be applied to a very wide range of systems, including atomic energy levels, particles in potential well, photons, and so on. Thus there is a lot of work to be done in disentangling the details of specific models of quantum mechanical systems from general features of quantum mechanics itself. 

One promising approach to this problem involves the use of operational theories\cite{CDK, PhysRevA.81.062348} and a related family of research programmes including nonlocal correlations\cite{Rohrlich, Pawlowski, Toner}, generalized probabilistic theories\cite{Hardyreasonable, Barrett, SpekkensBarrettLeifer, Masanes_2011}, empirical models\cite{Abramsky_2011}, and device-independent approaches\cite{newlabelColbeck, AdlamKent2, kthw,Acin}. These frameworks all locate quantum mechanics within a larger space of theories and then adopt a strategy of studying the relations between these theories as a way of characterising and understanding quantum mechanics itself.  Moreoever, all of these frameworks define the theories under consideration  in terms of observable quantities -  preparations and measurements for operational theories, correlations between observed events for nonlocal correlations, or measurements and contexts for empirical models. The hope is that by removing all the details of specific physical realisations we can study the structure of the resulting abstracted theory and thus come to a better understanding of what quantum mechanics is really about. 

The specific sort of structure that we identify in this operational context often takes the form of relations between various possible features of the operationally defined theories. In particular, it is common to see proofs of the form  `every theory which has property $A$ also has property $B$,' where usually there seems to be an implication that property $A$ is supposed to explain property $B$\cite{Pawlowski, Toner, Masanes_2006}. For example, Masanes, Acin and Gisin have shown that every theory which obeys a no-signalling constraint must also exhibit a monogamy bound limiting the strength of multipartite non-local correlations\cite{Masanes_2006}. 

These operational frameworks are also often used to come up with new operational axiomatisations of quantum mechanics. Operational axiomatisations  have a long history - even von Neumann himself came up with one\cite{hilbert1928grundlagen, birkhoff1936logic}. But the recent spate of interest in axiomatisations seems to have begun with Hardy's `Quantum Theory from Five Reasonable Axioms.'\cite{Hardyreasonable} Hardy derived the mathematical formalism of quantum mechanics from a set of axioms including the structure of composite systems and the existence of a continuous reversible transformation between any two pure states. Hardy notes that all of these five axioms are also true within classical physics, apart from the existence of the continuous reversible transformation, and even that can be made compatible with classical physics if we simply omit the word `continuous.' This one word, Hardy argues, is responsible for the difference between classical physics and quantum physics. A later example is 'Existence of an information unit as a postulate of quantum theory' by Masanes, Muller, Augusiak and Perez-Garcia\cite{Masanes_2013}. The main axiom used here is the titular `existence of an information unit' and this information unit is required to satisfy several further properties, including `no simultaneous encoding,' which states that if the information unit perfectly encodes a classical bit, it cannot also encode anything else. The derivation also requires the axioms of `continuous reversibility,' and `tomographic locality' and it is shown that quantum mechanics is the only theory satisfying all of these postulates.

Since these various operational approaches have many common elements, many of our comments will apply equally well to any of them. However, it is helpful to have a specific example in mind, and thus we will henceforth focus on operational theories.\footnote{`Operational theories' may be regarded as an umbrella term, with the other approaches we have mentioned representing some particular implementation of the operational approach. For example, generalized probabilistic theories are operational theories which are characterised in terms of state spaces, effect algebras and channels\cite{plavala2021general,}, while the term  `device-independent' typically refers to applications of operational theories within cryptography.} In this framework, theories are specified entirely in terms of preparation procedures and measurement results: a operational theory (OT) is a quadruple $(\mathcal{P},\mathcal{M}, \mathcal{T}, p)$ where $\mathcal{P}$ is a set of preparations, $\mathcal{M}$ is a set of measurements, $\mathcal{T}$ is a set of transformations, and the function $p(M^x | P, T) $ specifies, for every possible combination of  preparation $P$, transformation $T$, and measurement outcome $M^x$, the probability distribution over the possible outcomes of every possible prepare-transform-measure (PTM) scenario - where a PTM scenario is a process in which  a system is  prepared according to a preparation $P \in \mathcal{P}$, subjected to transformation $T \in \mathcal{T}$ (which could be the identity) and then some measurement $M \in \mathcal{M}$ is performed, yielding various possible measurement outcomes with the probabilities prescribed by the operational theory.\cite{spekkenscontextuality} Thus an operational theory amounts to specifying the outcome probabilities for every possible combination of operational procedures.  Note that in general the way in which preparations, transformations and measurements are labelled has no physical significance, so two sets  $(\mathcal{P},\mathcal{M}, \mathcal{T}, p)$,  $(\mathcal{P}',\mathcal{M}', \mathcal{T}', p)$ should be understood as defining the same OT if we can find an isomorphism from $\mathcal{P}$ to $\mathcal{P}'$, an isomorphism from $\mathcal{T}$ to $\mathcal{T}'$, and an isomorphism from $\mathcal{M}$ to $\mathcal{M}'$, such that the combination of these three isomorphisms takes $p $ to $p' $. We will henceforth take it that this quotienting procedure  is implicit in the definition of an OT, so we need not be concerned with the way in which the procedures are labelled.

\section{Axiomatisations \label{counterfactual}}

There are a number of good reasons for being interested in research in the operational theories framework. These results may be regarded  as a means of increasing our familiarity with the theory, or increasing our subjective feelings of understanding, or they may serve a pedagogical purpose. They also provide  us with a `device-independent' perspective \cite{AdlamKent2,kthw,Acin,HR,bckone} which allows us to provide proofs of security for cryptographic protocols depending only on simple principles like the no-signalling principle, which means  the security proofs hold even for participants who are not in a position to be certain about the exact makeup of the devices used in the protocol - for example, if they have to rely on devices made by a manufacturer who they don't trust. 

But on top of these pragmatic motivations, many people working on operational theories argue that their research has some sort of foundational significance  - it provides `\emph{insight into the reasons why quantum theory is the way it is}'\cite{Hardyreasonable}. Let us therefore turn to an examination of what sort of insight we might gain from this sort of research.  

It has been suggested that perhaps we should abandon the project of interpreting quantum mechanics altogether and instead attempt to understand quantum mechanics via axiomatisations\cite{1996IJTP...35.1637R, 10.1086/525620}. As a first response to this argument, let us reinforce that `interpreting' quantum mechanics is not merely about our desire to have a subjective feeling of understanding: for without an interpretation we don't have a precise criterion for when we should ascribe unitary evolution to systems and when we should instead ascribe non-linear collapse as in the quantum measurement process, and there are cases where making different choices on this point can lead to different empirical predictions\cite{Frauchiger_2018,adlam_2021}. In general axiomatisations of quantum mechanics  take measurement as an unanalysed primitive, so they offer no resolution to this particular scientific problem. The proponent of the axiomatic approach might perhaps try to make progress by taking measurement to be defined functionally by the axioms, and then if the axioms were detailed enough in the right sort of way it might perhaps be possible to offer a solution to the measurement problem along these lines, but to our knowledge no one has yet taken this route. Thus at present it seems that we can't simply replace interpretation with axiomatisation.

But let us put the measurement problem aside for now and meet the axiomatisation programme on its own terms. We will henceforth suppose that `preparations,' `transformations,' and `measurements,' are all perfectly well-defined categories; and we will argue that even in this simplified context, axiomatisation still can't entirely replace interpretation, because the process of axiomatisation is itself in need of interpretation. 

Schematically, an axiomatisation of quantum mechanics proceeds as follows. First, we  set out a number of axioms such as `no-signalling' or `continuous reversibility,' or `tomographic locality.' Each of these axioms defines a subset of the space of OTs - that is, the set of all OTs which obey the axiom. Next, we demonstrate that there is only one OT in the intersection of all of these sets, and that this OT  is the OT representation of quantum mechanics. In fact, because of the lack of a clear definition of `measurement' in conventional quantum mechanics, there may be some ambiguity about exactly how quantum mechanics should be expressed as an OT;  however, as we have agreed to ignore the measurement problem for the moment, we will also ignore this ambiguity and thus will henceforth suppose that there is a unique OT, QMOT  (Quantum Mechanical Operational Theory), which can be identified as the OT representation of quantum mechanics. That is to say, to axiomatize quantum mechanics is simply to find a way of choosing subsets of the space of OTs such that the intersection of these subsets contains only QMOT.

The problem with this approach is that there are a great many possible ways to come up with such an axiomatisation. Indeed, since there are an infinite number of possible OTs (owing to the continuity of probability functions), there are an infinite number of possible `axiomatisations.' For example, we can always select two random OTs $A$, $B$ and propose two extensionally defined axioms to the effect `the operational theory which holds in the actual world must belong to the set $\{ A, QMOT\}$' and `the operational theory which holds in the actual world must belong to the set $\{ B, QMOT\}$.'  These axioms clearly suffice to pick out QMOT uniquely as the OT which must hold in the actual world, but no one would consider that there is any insight or understanding to be gained from this `axiomatisation.' So clearly it is not the case that we can use just \emph{any} set of OTs for our axiomatisation; the sets are required to correspond to features which are interesting or relevant along some axis.

In ref \cite{Rovelli} Rovelli tells us that the axioms should have `clear physical content,' and should be known to be  `experimentally true,' but this does not seem to narrow things down very much - after all, an assertion like `the actual mosaic must belong to the set $\{ A, QMOT\}$' certainly makes a clear and unambiguous statement about physical reality, and insofar as we know that quantum mechanics is experimentally true, we know that this axiom is experimentally true. Likewise, Grinbaum suggests that the axioms `must be simple physical statements whose meaning is immediately, easily accessible to a scientist’s understanding,' but again, this doesn't seem to help a great deal, as `the  actual mosaic must belong to  the set $\{ A, QMOT\}$' is a statement about physical reality, and at least from a certain point of view it seems fairly simple and easy to understand. Evidently this is not the sort of simple physical content that Rovelli and Grinbaum have in mind, but this only reinforces the fact that we require some clarity on what they \emph{do} have in mind. Similar points can be made about other popular ways in which researchers have described their axioms: \emph{ `simple,'}\cite{Masanes_2011} \emph{`natural,'} \cite{Masanes_2011} \emph{ `elementary,'}\cite{CDK} or \emph{`reasonable,'}\cite{Hardyreasonable}.   

Moreover, it is sometimes suggested that OT research might help to shape future theories\cite{M_ller_2021} - the idea seems to be that if we can derive certain features of quantum mechanics from simple, foundational principles, then we can have a high degree of confidence that these features will survive future theory changes.  However, this argument runs up against a version of the pessimistic meta-induction: past scientific revolutions have often  overthrown `simple, foundational' principles that earlier scientists might well have had a great deal of confidence in, so we should perhaps be careful about assuming that our own favoured principles are going to survive. For example, prior to the formulation of Special Relativity many physicists presumably had a great deal of confidence in the `simple, natural, reasonable' idea that parallel rays of light could never intersect. Therefore it is important to give some account of why it is reasonable to believe that certain sorts of principles will persist through theory change: vague intuitions around how simple and foundational they look don't really cut it, not least because people's intuitions on these points differ substantially. So in order to draw sensible conclusions about how research in the OT framework might aid searches for future theories, we are in need of some account of how the process of axiomatisation is to be understood, and why we judge certain axioms and not others to be likely to survive theory change. 

Of course one might argue that axiomatisations are  chosen for purely practical reasons: it's useful to have a simple formulation of a theory in order to make the calculations easier.  But this doesn't seem to be what is going in in the case of the OT framework, as there is no sign of anyone trying to replace the mathematical methods of quantum mechanics with a formulation in terms of OTs, and nor does it seem likely that calculations in the OT framework would be any easier than the standard ones. This  practical motivation also would not explain the claim that such axiomatisations explain something about why quantum mechanics is the way it is, or the idea that they might be a guide to future theories.

A second possibility is that the preference for simple axiomatisations is to be understood from a Humean perspective in the spirit of Lewis' best-systems approach\cite{Lewis1986-LEWOTP-3, lewishumean}. In this picture, the laws of nature are the axioms that appear in the best systematisation of everything that actually happens in the universe; so as  good Humeans we are supposed to look for the best systematization of our existing data and then suppose that  this systematization corresponds  to the laws of nature as specified by the real best system. Different axiomatisations of quantum mechanics can therefore be regarded as competing proposals for best systems, and our preferences over axioms can be understood as a process of evaluating axioms against one another for simplicity, strength and fit in order to determine which one is most likely to correspond to the real best system. This is in some ways quite a satisfying account of the process of axiomatisation, and indeed, it provides a good counterexample to the assertion that the best-systems approach does not give a realistic picture of the way science really works\cite{pittphilsci14730}: here we have a real community of scientists apparently engaged in searching for a best-system. 

However, there are a number of objections to the best-systems approach which would also constitute objections to this interpretation of the axiomatisation programme\cite{doi:https://doi.org/10.1002/9781118398593.ch17}\footnote{For example, it has been argued that Humean laws can't licence inferences from the observed to the unobserved\cite{Berenstain2012-BEROSR}; that Humean laws make scientific explanations circular\cite{Maudlin2007-MAUTMW}; that the Humean picture  entails that the existence of regularities is an inexplicable miracle\cite{Strawson2014-STRTSC-5}; and that the Humean picture does not account for the important role of modal concepts in scientific discourse\cite{Berenstain2012-BEROSR}. The debate between Humeans and non-Humeans is ongoing, but these concerns seem reason enough to take non-Humean approaches seriously.} We won't have space here to pursue these objections, so let us henceforth simply take it for granted that we are not convinced by Humeanism and thus we are seeking a realist, non-Humean interpretation of the operational theories approach. In that case, a final possibility is that axiomatisations are to be regarded as \emph{hypotheses}: thus we prefer simpler axioms for just the same reasons as we usually prefer simple hypotheses. Moreover it is natural to suppose that a if hypothesis is correct, then it is likely to be a good guide for future theory-building.  But in order to substantiate this claim, it is necessary to give some account of what axiomatisations are supposed to be hypotheses about - that is to say, we need some sort of \emph{interpretation} of the operational theories framework in order to understand why certain sorts of axiomatisations are interesting. Let us therefore turn to the question of how one might choose to interpret the operational theories framework from a realist point of view.

\subsection{Structural Realism}  

Traditionally scientific realism has been formulated in terms of realism about theoretical entities: for example, Boyd lists as one of the central theses of  realism the idea that `\emph{theoretical   terms   in   scientific    theories   (i.e.,    nonobservational    terms)  should  be  thought  of  as  putatively  referring  expressions.}'\cite{Boyd+1984+41+82} And on the one hand, it is clear that operational theories are indeed concerned with entities, since the fundamental building blocks of an operational theory are local physical systems which can be combined with other local physical systems via a tensor product operation. But these entities are not \emph{theoretical} entities - they are usually understood as macroscopic `black boxes' which we can observe and interact with directly, sometimes putting inputs into them (e.g. by choosing a measurement direction) and sometimes extracting outcomes from them (e.g. by observing a measurement result). Moreover, because  the operational approach aims to abstract the underlying structure of quantum mechanics away from specific physical implementations, the systems in an operational theory are not endowed with any properties or internal structure. Thus being committed to an operational theory involves no commitment to any kind of theoretical entities - merely a commitment to what the theory says about observable phenomena, i.e. its empirical adequacy.  The operational theories framework thus appears to have much in common with \emph{constructive empiricism}, the prominent empiricist position developed by van Fraassen which insists that accepting a theory involves  nothing more than believing that it is empirically adequate\cite{van1980scientific}.

Conversely, from the point of view of traditional object-oriented realism the operational theory approach seems somewhat puzzling - for if one believes that science is all about describing the properties and behaviours of specific theoretical entities, then it's hard to understand what we're doing when we go about abstracting away the details of the specific physical systems we are dealing with. The traditional realist can regard quantum mechanics itself as a theory of atoms, particles in potential wells, photons and so on - but what is an operational approach to quantum mechanics supposed to be a theory \emph{of}? 

One promising option for committed realists is to move to a different form of realism - specifically, \emph{structural realism}.  Recall that  \emph{epistemic} structural realism is based on the idea that we can only find out about objects via relations between different objects, and thus we can never really find out about the essential nature of the objects in and of themselves\cite{Morganti2004}. Epistemic structural realism thus offers  a way of making sense of the abstraction process at the heart of the operational theories approach: if indeed it is the case that we can never really find out about objects in and of themselves, it makes sense to refrain from ascribing properties or features to the `systems' that we study, and instead define them entirely in terms of  the ways in which they relate to other objects  via accepting inputs and producing outputs, i.e. the role they play in operationally defined structures. 

Alternatively, \emph{ontological} structural realism in its strongest form says that there are no objects at all, only relations\cite{Ladyman1998-LADWIS-2}, whilst milder forms of ontic structural realism suggest that objects may exist but relations are ontologically primitive\cite{Ainsworth2010-AINWIO-2}, or that objects have their identities only in virtue of the relations in which they stand\cite{pittphilsci5531}. This idea offers a slightly different way of making sense of the abstraction process in the operational theories framework: if indeed it is the case that the ultimate stuff of reality is fundamentally relational, then the `systems' featured in an operational theory are not really objects at all, they are merely placeholders, i.e. nodes of relations. So naturally we should not ascribe properties or features to them, and therefore it is entirely appropriate to define them entirely via the operationally defined relations in which they stand. This is obviously a very different picture of the world from the object-oriented story centered around physical systems that the operational theories framework appears to present us with, but since the OT framework does not attribute any non-relational features to these physical systems, there is actually no obstacle to getting rid of objects entirely and regarding them as simply `nodes of relations,' and thus despite the apparent focus on objects this is actually a very natural target for an ontic structural realist approach. 

  In this article we will make no attempt to adjudicate between the epistemic and ontological approaches: rather we will focus on elucidating what \emph{kind} of structure we might be dealing with, and what we might have to learn from studying such structure. We reinforce that it is not our intention to argue that all researchers in the field actually think of their research in this way, or that this is the only possible way to interpret their work;  indeed, as noted by Letetre in ref \cite{Letertre2021-LETTOF}, `\emph{the realist/antirealist debate is purely epistemic and does not involve any feature of any formalism ... (and therefore) the operational framework for physical theories in itself does not favour a particular epistemic or ontological stance.'} Thus we don't claim that the operational framework in any sense \emph{mandates} a realist or structural realist approach. Rather, we seek to understand how one might interpret the results of the operational theories research programme if one comes to the study of operational theories having already adopted scientific realism for independent reasons.

\subsection{Modal Structural Realism}

An additional challenge to a realist take on  research in the OT framework is that many results in the field are concerned with \emph{counterfactuals}, since we are dealing with a large space of possible OTs, only one of which holds in our actual world. One might therefore worry that there is no possible realist interpretation of this sort of research, since the entities to which  the results refer are largely not real.  However, it has been observed\cite{10.1093/bjps/axl020} that if structural realists are  to avoid the Newman problem, they would do well to postulate structures which include \emph{intensional relations}, and many proponents of structural realism have argued that these relations will have to be \emph{modal} relations\cite{ladyman2007every,esfeld2009modal} - for example, causal relations, counterfactual dependence, or metaphysical necessitation. Moreover, these sorts of modal relations are frequently analysed in the possible world semantics, which employs a space of possible worlds and defines modal relations in terms of relations between sets of worlds\cite{https://doi.org/10.1002/malq.19630090502}. There is an obvious analogy to be made between the space of possible worlds used in modal logic and the space of possible theories invoked by the OT framework, which suggests that research in the OT framework may be understood as a way of  studying modal structure within a structural realist approach. In fact, we will argue that much of this research  can  be interpreted in terms of  modal structure defined by the relation of metaphysical necessitation. 
 
 \subsection{Selective vs Non-Selective Realism }

 \emph{Selective realism} is a form of scientific realism where realist commitments are carefully limited to specially selected features of the theory, such as working posits or unobservable entities, in the hope of avoiding the threat of the pessimistic meta-induction\cite{Papineau1996-PAPTPO-3}. Structural realism is sometimes regarded as a form of selective realism. But as argued in ref \cite{Ladymanstruc}, one may also pursue a non-selective approach to structural realism: we can maintain that the success of past theories is explained by continuity of structure through theory change without  being committed to the view that we can always identify \emph{which} specific structures will survive \emph{future} theory change. So we have a choice to make about whether we should adopt a selective strategy or a non-selective strategy for our structural realist interpretation of the operational theories framework. The former option would entail that different axiomatisations in the operational theories approach should be regarded merely as \emph{redescriptions} of the same underlying reality, while the latter would entail that different axiomatisations should be regarded as \emph{competing hypotheses} which represent different possible modal structures for reality. 
 
Of course, inclinations on this point are likely to depend to some degree on pre-existing metaphysical commitments. For example, suppose you have a fairly robust realist attitude towards modality - you believe the world has some unique, well-defined and objective modal structure (as do the authors of refs \cite{Berenstain2012-BEROSR, Ladyman2007-LADETM, esfeld2009modal}, amongst others). As we will argue in section \ref{mn}, axiomatisations of quantum mechanics can certainly be regarded as expressing some sort of modal structure, and therefore if you believe that reality has a unique objective modal structure, you are also likely to believe that different axiomatisations could potentially differ in how much they get right about that objective modal structure. On the other hand, if you don't believe that modal structure is unique or well-defined then you may be more inclined to pursue the selective realist strategy and insist that different axiomatisations are all equally valid redescriptions of the same reality. But the latter approach leaves us with what appear to be fairly weak realist commitments - if every different modal structure we can come up with for quantum mechanics is equally correct, then it seems that the `structure' we are supposed to be adopting an attitude of realism towards is merely some sort of vague `modal mush,' which doesn't seem likely to engender better understanding or theoretical progress. This approach also seems somewhat at odds with the language used by researchers in the field, who often suggest that their derivations have `explained' ... `shown why' etc. Of course one might take the view that there are many different equally valid `reasons why' quantum mechanics is the way it is, but if the `reasons why' are not unique it seems less clear why we should be interested in them.
 
Moreover, in the words of Pooley, sometimes \emph{`for advances to take place, it is important that the different formulations are considered to be genuinely distinct and exclusive alternatives.'}\cite{Pooley2005-POOPPA} For example, in a sense Einstein's seminal 1905 paper\cite{Einstein1905} was just a `redescription' of the existing Lorentz transformation equations, but it was Einstein's description and not Poincar\'{e}'s which led to the scientific revolution of special relativity; and most people would probably be inclined to say that the reason Einstein's approach led to progress was because Einstein was \emph{right} (or at least, more right), i.e.  the modal structure he proposed, where the Lorentz transformation equations follow from the relativity of simultaneity, gave more insight into the actual modal structure of reality than Poincare's approach where they are simply postulated as a dynamical feature of electromagnetic systems\cite{poincare2003science,adlam2011poincare}. Of course, this is not to say that we should \emph{always} consider reformulations of a theory to be genuinely distinct. But presumably we do want to consider formulations distinct when they point to different possible routes for scientific progress, and since it's  at least possible that different axiomatisations could suggest different research programmes, the non-selective approach to axiomatisation should be taken seriously. Thus we will henceforth  suppose that different axiomatisations may sometimes represent different possible modal structures; the task of the remainder of the paper will therefore be to elucidate what sort of modal structure they could represent, and what we might have to learn from them.

\section{Structure \label{structure}}

\subsection{Metaphysical Necessitation \label{mn}} 

Recall that the `Humean mosaic' of a possible world contains all the actual, non-modal facts in that world. Humean mosaics are often thought of as distributions of categorical properties across spacetime, but here we do not intend to be prescriptive about the content of the Humean mosaic - different scientific theories may make very different suggestions about what  `actual, non-modal' facts consist of, so to avoid committing to any one theory we will avoid specifying what the Humean mosaic should look like.

Let us suppose that we are given some Humean mosaic and we wish to map it to an OT. First of all, observe that this may not always be possible - if a given mosaic does not exhibit enough regularities, or does  not contain agents who perform preparations/measurements/transformations, there may be no sensible way to associate an OT with it. Let us therefore define a special subset of `OT-mosaics' such that it makes sense to associate the mosaic with an operational theory: 

\begin{enumerate} 
	
	\item OT-mosaics contain local regions  which can unambiguously be identified as `preparations,' `transformations' and `measurements,' and there is some natural way of identifying repeated instances of the same preparation, transformation or measurement  
	
	\item OT-mosaics exhibit a high degree of homogeneity - i.e. within any randomly selected spatiotemporal region of reasonable size, the relative frequencies exhibited within that region are close to the relative frequencies across all of spacetime

	\end{enumerate} 

The first point is necessary in order that there is something for the sets $ \mathcal{P},\mathcal{M}, \mathcal{T}$ defined by the OT to refer to, and the second point is necessary so that regularities observed in one region of the mosaic can be taken as reflective of regularities across the whole mosaic.  

Because  there is a natural way to identify the preparations, transformations and measurements in an OT-mosaic, after making that identification we can disregard the other contents of the mosaic and simply regard an OT-mosaic as a very large collection of PTM scenarios. In order to regard this collection of PTM scenarios as the model of an OT we still need to define the probability distributions $p$ that the  OT assigns to the PTMs. Let us first consider the simplest case, where every possible PTM scenario (i.e. covering every possible combination of   $\mathcal{P},\mathcal{M}, \mathcal{T}$) actually occurs; then we can simply stipulate that the distributions assign probabilities which are equal to the relative frequencies exhibited by the actual outcomes obtained in this particular Humean mosaic.  Thus, using this prescription, we can map an OT-mosaic to a unique OT. We will refer to this mapping as the OT-map.
 
Therefore the  relation $R(A,B)$ expressed by the statement `every OT which has feature $A$ also has feature $B$'  can be rewritten in the form `every possible world containing an OT-mosaic which is mapped by the OT-map to an operational theory having feature $A$ is also a possible world containing an OT-mosaic which is mapped by the OT-map to an operational theory having feature $B$.' This should start to look familiar, since the relation of metaphysical necessitation is usually analysed in the possible world semantics such that `$X$ metaphysically necessitates $Y$' if and only if `every possible world in which $X$ is true is also a world in which $Y$ is true.'  Thus $R(A, B)$ can be understood as expressing the fact that conditional on our being in an OT-mosaic, feature $A$ metaphysically necessitates feature $B$. That is to say, a piece of research demonstrating that some feature $B$ can be derived from some other feature $A$ can be understood as making a statement about the \emph{real} modal relation of metaphysical necessitation, rather than about some space of counterfactual theories. 

However, note that in most realistic cases we will not typically be able to define an OT-map in this straightforward way: in general stipulating that the probabilities match the actual relative frequencies will not give a well-defined   operational theory, as there may be some PTM scenarios which do not occur at all in the mosaic. Indeed, given that QMOT defines an infinite number of PTM scenarios, no finite-sized mosaic can possibly be mapped to it by the OT-map (and therefore it's likely that the actual Humean mosaic is not taken to QMOT by the OT-map). Moreover, we also want to allow for the possibility that the relative frequencies exhibited in a mosaic may be close to the   values in the underlying probabilistic theory but not an exact match. So let us also define an `approximate OT-map' which takes each OT-mosaic to one or more operational theories such that the  mosaic is `consistent' with all of these operational theories. Here, we say that   a mosaic is consistent with an OT if the probability assigned to that mosaic by the OT is larger than some threshold value - the exact choice of threshold value will be relative to one's preferred standards for empirical confirmation. The law of large numbers entails that whenever a mosaic contains a large number of repetitions of some particular PTM scenario, if the relative frequencies exhibited in the outcomes for these scenarios are significantly different from the probabilities prescribed by an OT, then the probability assigned to the mosaic by that OT will be very close to zero;  so if the OT-mosaic is  sufficiently large, for most reasonable choices of threshold value only a very small subset of the possible operational theories will be considered consistent with the mosaic. For example, it is generally agreed that QMOT is more or less the only operational theory which is consistent with our observations in regimes relevant to quantum mechanics. 

  Thus in the more realistic case,   the  relation $R(A,B)$ expressed by the statement `every OT which has feature $A$ also has feature $B$,'  can be rewritten in the form `every possible world containing an OT-mosaic which is mapped by the approximate OT-map to (at least one) operational theory having feature $A$ is also a possible world containing an OT-mosaic which is mapped by the approximate OT-map to (at least one) operational theory having feature $B$.'  Thus $R(A, B)$ can be understood as expressing the fact that conditional on our being in an OT-mosaic, the fact that our mosaic is consistent with an OT having feature $A$ metaphysically necessitates that our mosaic is consistent with an OT having feature $B$. That is to say, the existence of this relation explains why seeing behaviour which looks highly consistent with feature $A$ should lead us to expect that we will also see behaviour which looks highly consistent with feature $B$. 

Since axioms likewise can be written as sets of operational theories (i.e. an axiom can be given an extensional definition as the set of operational theories obeying the axiom), an axiomatisation of quantum mechanics is simply a special case of the relation $R(A, B)$, where $A$ is a conjunction of several axioms and $B$ uniquely singles out QMOT. Thus an axiomatisation of quantum mechanics can likewise be understood as expressing a modal fact: the axioms are a set of constraints which are together sufficient to metaphysically necessitate the operational theory QMOT, and thus an axiomatisation can be regarded as telling us that it is metaphysically necessary that if we see behaviour which looks highly consistent with all of the axioms separately, then we will  also see behaviour which looks highly consistent with QMOT.

\subsection{Laws of Nature \label{GL}}

\subsubsection{Constraints} 
We have identified a way in which operational theories can be associated with a form of modal structure. So far, what we have said would be compatible with the selective realist view that all axiomatisations are equally valid - and those who prefer this approach are of course free  to make use of the ideas developed in the previous section without accepting the rest of the analysis. But we have argued for the non-selective approach where we suppose that reality has some objective modal structure which a given axiomatisation may or may not successfully capture, and thus we will now proceed to suggest how to reconcile axiomatisations of quantum mechanics with this non-selective worldview. 

Henceforth we will define the `objective modal structure' of reality in terms of laws of nature, in accordance with the view that laws of nature can be regarded as a form of objective modal structure, which is advocated in refs \cite{Berenstain2012-BEROSR, Ladyman2007-LADETM, esfeld2009modal} amongst others. In order to formalise this modal structure, we will employ the framework developed in  \cite{adlam2021laws, adlam2021determinism},  which characterises laws of nature in modal terms as \emph{constraints}. This approach is inspired by the increasing prominence of constraint-based laws in physics, as  surveyed in  \cite{adlam2021laws, adlam2021determinism}; a very similar account of lawhood in terms of constraints was also recently given by  \cite{chen2021governing}. To characterise constraints we again make use of the notion of a Humean mosaic. We emphasize that the use of this terminology is not intended to reflect a commitment to the standard Humean ontology consisting only of the Humean mosaic -  we use the phrase `Humean mosaic' to refer to all of the actual, non-modal content of reality, but in addition to the Humean mosaic we are committed to the existence of modal structures which are distinct from the Humean mosaic. In particular, we maintain that a `constraint' is something which is distinct from the actual Humean mosaic, but  which also plays some role in determining the content of the Humean mosaic, and therefore constraints must be regarded as  \emph{ontologically prior} to the Humean mosaic. 

In order to make this  account more concrete, we will henceforth define constraints \emph{extensionally}, appealing to techniques employed in modal logic: a constraint will simply be defined as a set of Humean mosaics, i.e. the set of all mosaics in which that constraint is satisfied. In some cases, a constraint will be expressible in simple English as a requirement like `no process can send information faster than light,' such that the  constraint corresponds to exactly the set of mosaics in which the requirement is satisfied. But there are many sets of mosaics which will not have any straightforward English characterisation, so we will not be able to state the corresponding constraint in simple terms. Nonetheless, each set still defines a unique constraint.

Using these definitions, we postulate that every world has some set of laws of nature which are an objective fact about the modal structure of that world, and we characterise these laws in terms of \emph{probability distributions over constraints}.  For example,  a deterministic evolution law which applies to systems of type $s$ is analysed as a probability distribution which assigns probability $1$ to the constraint consisting of the set of all Humean mosaics in which all systems of type $s$ have histories which are dynamically possible according to the law. Similarly, given an indeterministic time evolution law, we can use the law to construct a probability distribution $p_s(h)$ over each possible history $h$ of the system $s$ and then analyse the law as a probability distribution assigning  probability $p_s(h)$ to the set of all mosaics in which system $s$ has history $h$, and so on for every system of the same type across the whole Humean mosaic. And a law which prohibits superluminal signalling can be analysed as a probability distribution which assigns a high probability to the constraint consisting of all the Humean mosaics in which no superluminal signalling occurs, i.e. in which every potential signalling process exhibits a correlation of zero across all instances in the mosaic.   In this picture, we can think of a constraint being drawn for each law according to the probability distribution induced by the law, and then the actual mosaic must lie in the intersection of all of the selected constraints.  That is to say, we are to think of laws `governing' by selecting whole sets of mosaics at once and thus narrowing down the set of physical possibilities for the actual mosaic. 

\subsubsection{Axiomatisations \label{axiom}}

To illustrate the relationship between laws of nature and axiomatisations, we will first consider a simplified case, where we limit ourselves to considering only mosaics which are taken by the (exact) OT-map to a well-defined operational theory, and we suppose that the laws of nature are all associated with trivial probability distributions, i.e. each of them simply assigns probability $1$ to a single constraint. Thus the laws of nature give rise to a set of constraints $\{ C \}$ and the Humean mosaic of the actual world must lie in the intersection of all of those constraints. It seems likely that the Humean mosaic of the actual world is an OT-mosaic, and therefore we may suppose that each of the constraints in  $\{ C \}$ has some support on the set of OT-mosaics. Thus from each constraint $C \in \{ C \}$ we can obtain a corresponding constraint $C'$ consisting of the support of $C$ on the set of OT-mosaics, so we obtain a  set $\{ C' \}$ of constraints with support only on the OT-mosaics. We are considering only OT-mosaics which are taken by the OT-map to a unique operational theory, and therefore to each of the constraints  $C' \in \{ C' \}$ we can associate a set of operational theories $O_{C'}$. And each set $O_{C'}$ can be regarded as extensionally defining an `axiom,' (i.e. we understand $O_{C'}$ to be the set of all operational theories obeying the axiom). For example, a  constraint consisting of the set of all Humean mosaics which  don't contain any signalling processes will give rise to a constraint consisting of the set of all OT-mosaics which are taken by the OT-map to operational theories which do not allow signalling - so this constraint leads us to a set of OTs which constitute the extensional definition of an axiom that forbids signalling. 

We now suppose  that the actual Humean mosaic lies in the intersection $I_c$ of all of the constraints $\{ C\}$, and that it is taken by the OT-map to QMOT. It follows that QMOT must lie in the intersection of all  of the sets of operational theories $\{  O_{C'} \}$.  That is to say, the laws of nature induce what is essentially a natural `axiomatisation,' because the constraints give rise to a set $\{  O_{C'} \}$ consisting of sets of operational theories whose intersection contains QMOT. Moreover, given some particular axiomatisation of quantum mechanics we can always reverse this procedure, obtaining from each axiom the constraint consisting of all OT-mosaics which are taken by the OT-map to OTs which obey this axiom, and thus we can think of any axiomatisation as a set of constraints. An axiomatisation of quantum mechanics can therefore be understood as proposing  a set of constraints which are to be regarded as possible candidates for the constraints $\{ C'\}$. And since at most one axiomatisation corresponds to the \emph{actual} set $\{ C'\}$ that follows from the laws of nature,  different axiomatisations may be regarded as competing hypotheses about the way in which quantum mechanics arises from the objective modal structure of reality. Note that nothing we have said thus far implies that the intersection of the sets $\{  O_{C'} \}$ contains \emph{only} QMOT, but we are of course free to hypothesize that, as a matter of fact, the `axiomatisation' encoded by the set $\{  O_{C'} \}$ singles out QMOT uniquely, as most  extant attempts to axiomatize quantum mechanics do.

Now, this case is clearly an idealisation - for as noted above, the actual Humean mosaic is probably not in fact taken by the OT-map to QMOT. Moreover, we may not be inclined to say that every law of nature assigns probability $1$ to a single constraint. For example, in the no-signalling case, we probably do not want to insist that the actual mosaic must belong to the constraint consisting of mosaics in which every potential signalling process exhibits a correlation of \emph{exactly} zero across all instances: since operational theories are probabilistic theories, we want to allow for the possibility that the relative frequencies exhibited in the actual mosaic diverge to some small extent from the expected ones. So instead we should allow that the laws actually correspond to nontrivial probability distributions over constraints. That is to say, for each of the constraints $C \in \{ C\}$, we should obtain a probabilistic version of the corresponding law which assigns probability very close to $1$ to a constraint $D$  such that every mosaic $M$ in $D$ is very similar to a mosaic $M'$ in $C$ - where by `very similar' we mean that the two mosaics $M, M'$ will be taken by the approximate OT-map to two sets of OTs which coincide except possibly on a set of OTs of measure zero\footnote{An operational theory can be written as a vector in $\mathbb{R}^n$ for some large or perhaps infinite $n$: for each PTM scenario we construct a vector containing the probabilities assigned by the OT to each of the possible measurement outcomes in a specified order, and then we can concatenate the vectors for each PTM scenario to give a vector in $\mathbb{R}^n$. Thus when $n$ is finite we can employ the Lebesgue measure over the set of operational theories; when $n$ is infinite (e.g. when we want to allow an infinite number of different possible PTM scenarios) we would have to use some other measure, such as the  Wiener measure. Thus we can always make sense of the notion  of a `set of measure zero' of OTs.}. Evidently $C$ is then a subset of $D$. For example, in the no-signalling case, we would propose a law which assigns probability very close to $1$ to a constraint $D$ consisting of  mosaics in which every potential signalling process exhibits a correlation of less than or equal to $\epsilon$ across all instances in the mosaic, where $\epsilon$ is  very close to zero. 

It then follows that the set of mosaics lying in the intersection $I_D$ of all of the constraints $\{ D\}$ is such that  every mosaic $M$ in $I_D$ is very similar to a mosaic $M'$ in $I_C$, where `very similar' is again understood relative to the approximate OT-map. Recall that all of the mosaics in $I_C$ are taken by the exact OT-map to operational theories in the intersection of $\{  O_{C'} \}$; so  all of the mosaics in $I_D$ except a set of measure zero will be taken by the \emph{approximate} OT-map to at least one operational theory in the intersection of $\{ O_{C'} \}$. If we suppose that the intersection of $\{ O_{C'} \}$ in fact contains only QMOT, it follows that all of the mosaics lying in the intersection $I_D$, except a set of measure zero, are taken by the approximate OT-map to QMOT (and even the mosaics in the set of measure zero will be taken by the approximate OT-map to at least one OT which is quite similar to QMOT). Thus when we draw constraints according to the distributions assigned by the probabilisitic laws, with very high probability we will end up with the set of constraints $\{ D\}$ and thus the actual mosaic, drawn from the intersection  $I_D$, will almost certainly  be taken by the approximate OT-map to QMOT, which means that it will exhibit behaviour which is highly consistent with QMOT. 

 In this way, the probabilistic construction still allows us to say that the modal structure captured by the  axioms $\{  O_{C'} \}$ explains  why our observations of quantum reality have yielded results showing relative frequencies which are extremely close to the predictions of QMOT:  just as the deterministic laws in the idealised case ensure that the actual mosaic will be taken by the OT-map to an OT which obeys all the axioms in $\{  O_{C'} \}$, thus metaphysically necessitating that the actual mosaic should be taken by the OT-map to QMOT, so the probabilistic laws in the more realistic case ensure that \emph{with high probability} the actual mosaic will be taken by the \emph{approximate} OT-map to an OT which obeys all the axioms in $\{  O_{C'} \}$, thus metaphysically necessitating that the actual mosaic should be taken by the \emph{approximate} OT-map to QMOT.  Therefore these axioms $\{  O_{C'} \}$ still encode the way in which the underlying modal structure gives rise to the observed quantum-mechanical behaviour, even though the axioms are most likely not \emph{exactly} obeyed by the relative frequencies in the actual Humean mosaic. And therefore we may still regard different axiomatisations  as competing hypotheses about the way in which  the operational behaviour associated with quantum mechanics arises  from the objective modal structure of reality.

 \subsection{Consequences} 
   
This interpretation of the axiomatisation programme has a number of useful consequences. First, it gives a straightforward explanation for the nature of the `understanding' which we gain from an axiomatisation of quantum mechanics - it is simply the usual sort of understanding which one gains from learning the reason or cause of something. Of course, we don't know for sure that any particular axiomatisation is in fact the correct one, but this is consistent with the way understanding works in more familiar cases: when someone suggests a possible reason for an event, if we find the explanation plausible we will usually take ourselves to have gained understanding, even if we have not been offered a watertight proof that the proffered reason is in fact the true reason.

 Second, this account explains our preferences for certain sorts of axiomatisations - if we take ourselves to be looking for the axioms $\{  O_{C'} \}$ which correspond to the specific set of constraints  $\{ C' \}$ which are  induced by the laws of nature, then our beliefs about the laws of nature will naturally inform our preferences over axiomatisations.  For example, if we believe that the laws of nature must be simple, intuitive or reasonable, then we will naturally expect that the set of operational theories corresponding to the constraints $\{ C' \}$ will be unified by some relatively simple, intuitive or reasonable feature, and thus the axioms that they define will be simple, intuitive or reasonable.  So there is no longer any mystery about the nature of the intuitions that induce us to favour  axiomatisations containing principles that look `simple,' `natural,' `elementary,' `reasonable' and so on: they are simply intuitions about what the laws of nature look like. 

And third, this account allows us to explain why axiomatising quantum mechanics is a worthwhile activity from the point of view of a practising scientist: for if we do manage to come up with an axiomatisation corresponding to the actual constraints $\{ C' \}$ induced by the laws of nature, we may be able to make backwards inferences about the laws of nature themselves, and such inferences are likely to be a useful guide when we are applying quantum mechanics to new domains or trying to unify it with other theories as in quantum gravity. For example, if we believe that the set $\{ C' \}$ includes a constraint $C'$ singling out the set of no-signalling OTs, by extrapolation we might infer that the laws of nature include a constraint $C$ which picks out the set of Humean mosaics which don't contain any superluminal signalling. As a matter of fact we already have independent evidence to this effect from relativity\cite{Einstein1905}, but if we had not already known that we might have been able to infer it from an axiomatisation of quantum mechanics which includes something like `no-signalling' as an axiom. Obviously this process is not algorithmic; we don't know for sure which axiomatisation corresponds to the actual set $\{ C' \}$, and even if we did there would be many possible sets $\{ C \}$ from which this set $\{ C' \}$ with support only on the OT-mosaics could be derived, but nonetheless when we seek a consistent, unified set of laws of nature from which the set $\{ C' \}$ could be derived, we will certainly uncover possible routes for new research. So the realist account of axiomatisations seems to do better than the anti-realist or Humean accounts in that it allows us to do justice to the intuition that research in the OT framework could usefully inform the search for successor theories. 

   \subsection{ Possible objections \label{possibility2}}

 At this juncture, one might worry that this approach seems to suggest that only the unique  axiomatisation which corresponds to the constraints $\{ C' \}$ induced by the actual laws of nature can give us real insight into quantum mechanics, which would entail that the insight apparently yielded by other axiomatisations is only illusory. However, this is not the case. Although on this view there is a unique correct axiomatisation, there are still many axiomatisations which are `close' to the correct one - for example, a proper subset of the axioms  might correspond to constraints in $\{ C' \}$,  or an axiomatisation could correspond to a combination of two constraints  in $\{ C' \}$, or it might split a single constraint from $\{ C' \}$ into several axioms, or it could be the case that its axioms are derivable in a straightforward way from the axioms extensionally defined by the constraints in $\{ C' \}$. Thus a variety of different axiomatisations may offer us insight by revealing different facets of the actual constraints $\{ C' \}$, even if none of them gets  $\{ C' \}$ completely correct. 

Another possible concern relates to the fact that in the more realistic construction, the modal structure can only yield the conclusion that the actual mosaic will with very high probability exhibit relative frequencies very close to those predicted by QMOT, so how can we be sure that the relative frequencies we have observed in the actual world even approximately match the ones predicted by whatever the real modal structure is? Well, of course we cannot; that is the inevitable consequence of any approach which combines realism about modality with a commitment to the existence of probabilistic laws. Anyone hoping to explain behaviour in the actual world by appeal to an intrinsically probabilistic theory must assume that observed relative frequencies at least approximately match the probabilities prescribed by the theory, and thus since our aim is to provide an epistemic reconstruction of the OT framework, we are also entitled to make use of such an assumption. There are in fact ways to avoid probabilistic laws in the constraint framework -  we could choose to  postulate  `frequency constraints,' which roughly speaking translate any law of the form `events of type A have outcome A with probability 0.8' to a constraint of the form `the set of mosaics in which (very close to) eighty percent of events of type E have the outcome A,' so that the fact that the relative frequencies match those predicted by the laws is itself lawlike. (See refs \cite{adlam2021determinism,Roberts2009-ROBLAF}, for further discussion of this approach). However, this involves a fairly dramatic revision of our usual ideas about probability, so for the purposes of the current project we will leave open the possibility that the laws are genuinely probabilistic, and thus that the matching between QMOT and the actual mosaic is only probabilistic and approximate.   

A further concern is that  quantum mechanics is not a complete and final theory of reality, and if we later discover special circumstances in which the predictions of quantum mechanics are incorrect, it may turn out that in fact the intersection of the constraints induced by the actual laws of nature does \emph{not} include an OT-mosaic which is taken by the OT-map to QMOT, even in the idealised version of the construction.  However, this doesn't mean that the project of axiomatisation is worthless. Even if the operational theory QMOT is not strictly speaking the right one, it is likely to be the case that there are features of the actual objective modal structure of reality which explain why it is that so much of reality appears to conform to QMOT, and thus by attempting to axiomatise quantum mechanics in this way we would still have a good chance of making useful discoveries about the underlying modal structure of reality. For example, it might be the case that quantum mechanics works in general in the  actual mosaic because it is possible to derive QMOT from a set of axioms $A$ such that all but one of the axioms in $A$ corresponds to a constraint featuring in the actual set $\{ C' \}$ so if we find $A$ and then ask questions about what happens when we relax one or more of the axioms, we still have a useful route to scientific progress.

 \subsection{Example: Existence of an information unit as a postulate of quantum theory \label{example}}
 
 To make our discussion more concrete, we now apply this framework to a specific axiomatisation of quantum theory. The axiomatisation of ref \cite{Masanes_2013} employs three axioms: 
 
 \begin{enumerate} 
 
 \item  \emph{Existence of an information unit}: there is a type of system (the generalized bit or gbit) such that the state of any other system can be reversibly encoded in a sufficient number of gbits, such that: 
 
 \begin{itemize} \item The state of a gbit can be characterized with a finite number of measurements
 	
 	\item All linear functions from the state space of gbits to $[0,1]$ correspond to outcomes of measurements that can in principle be performed
 	
 	\item The group of time-continuous reversible transformations for two gbits contains at least one element which is not the product of individual transformations on the two gbits, so gbits can interact

 \end{itemize} 
 \item \emph{Continuous reversibility}: for every pair of pure states one can in principle engineer a time-continous reversible dynamics which brings one state to the other 
 
 \item \emph{Tomographic Locality}: the state of a composite system is completely characterized by the correlations of measurements on the individual components 
 
 \end{enumerate} 
 
One way to interpret this axiomatisation would be to regard it as expressing the modal structure induced by four laws of nature of roughly the following form: 

\begin{enumerate}

	 \item A law which assigns probability $1$ to the constraint consisting of all Humean mosaics which instantiate only one type of information

	 \item The absence of any law which assigns probability $0$ to the constraint consisting of all Humean mosaics in which a particular continuous transformation between categorical properties is instantiated  	  
	  
	\item The absence of any law which assigns probability $0$ to the constraint consisting of Humean mosaics in which a particular categorical property is detected in individual measurements on the parts of the system  
	
	 \item A law which assigns probability $1$ to the constraint consisting of all OT-mosaics 
	\end{enumerate} 

These putative laws are by no means the unique possible way of mapping axioms 1) - 3) to laws; they are also not very precise,  because we don't know exactly what sorts  of things feature  in the set of Humean mosaics to which they pertain.  But they do illustrate broadly the way in which general laws could reduce down to axioms within the constraint framework: the idea is that laws 1) - 4) are together sufficient to metaphysically necessitate quantum mechanics, and therefore the axiomatisation of ref \cite{Masanes_2013} can be read as a \emph{hypothesis} about the way in which the objective modal structure of reality gives rise to quantum mechanics. Note that here we have suggested deterministic laws which define exact constraints, but as discussed in section \ref{axiom}, in a realistic case we would expect to be dealing with probabilistic versions of these constraints. For example, what it is to `instantiate only one type of information' is expressed by ref \cite{Masanes_2013}  in terms of requirements on the probability distributions prescribed by an operational theory, so it is to be expected that in the real world the observed relative frequencies may not meet these requirements \emph{exactly}, though one would expect them to come very close.  Therefore law 1) might become something like `A law which assigns very high probability to the constraint consisting of all Humean mosaics which instantiate only one type of information, except in a few extremely rare instances.'  

Axioms 2) and 3) a  have been analysed not as laws but as the absence of certain laws. This will generally be the case  where we are dealing with axioms concerning possibility, because of course we don't want to analyse an axiom saying that some process is  possible as a law which says that process must actually be instantiated. Yet in order to analyse this axiomatisation within the constraint framework we still need to be able to postulate  that it is a lawlike matter that all these transformations  are possible, and thus we   analyse `all transformations between pure states are possible' as the absence of any laws forbidding transformations between pure states. Similarly, we have analysed `there are no properties which show up in joint measurements on composite systems but not in individual measurements on parts of the system' in terms of the absence of any law forbidding a categorical property from being detected in individual measurements on the parts of a system -  it then follows that any categorical property which is detectable at all can in principle be detected in individual measurements on the parts of a system, in accordance with \emph{tomographic locality.}

Moreover, we can see from this example the way in which hypotheses about modal structure might inform the search for new theories. For example, it is shown in  ref \cite{Masanes_2013} that QMOT obeys axiom 1) -  i.e. every OT which is QMOT is also an OT in which there exists a unit of information in the above sense. Thus mathematically speaking, we can derive `This OT exhibits a unit of information' from `This OT is QMOT.' But also, as shown in ref \cite{Masanes_2013}, every OT which obeys axioms 1), 2) and 3) is QMOT, so we can also derive `This OT is QMOT' from `This OT exhibits a unit of information' together with the other two axioms. Mathematically, the derivation can go either way: we could have quantum mechanics because of the existence of a unit of information, or we could have a unit of information because of quantum mechanics! So the \emph{hypothesis} the actual world is governed by  quantum mechanics because  a unit of information exists in the real world is a claim about \emph{modal} structure which goes beyond the purely mathematical facts. And this hypothesis has real scientific consequences.  For if we take it that systems of a certain sort possess a unit of information in the above sense precisely because they are governed by the laws of quantum mechanics, then we have no reason to expect that other types of systems or successor theories to quantum mechanics will satisfy the requirement that they should allow for the existence of a unit of information. But if we take it that systems of a certain sort obey the laws of quantum mechanics \emph{precisely because} there is a deeper, more fundamental requirement which looks something like law 1), then we have good reason to suppose that \emph{other} sorts of systems and successor theories should also exhibit a unit of information, which may serve as a useful way to narrow the search space for future theories.  Taking these alternative directions of explanation seriously and hypothesizing that the second direction is true therefore provides us with a possible route to scientific progress, whereas merely viewing these axiomatisations as redescriptions does not have this nice consequence: if axiomatisations aren't regarded as hypotheses about the real modal structure of the world, there is much less reason to think they could provide useful constraints on future theories. 

\section{Implications \label{cosq}} 

\subsection{Structural Realism \label{SR}} 

 Thus far we have focused on using structural realism to provide an interpretation of the operational theories approach. But conversely, we also submit that the operational theories approach may offer valuable insights to the structural realist programme, for it offers a mathematically sophisticated framework in which various possible modal structures can be formulated and the relationships between them can be studied. This is a tool which has so far been lacking from structural realist approaches - typically the `structures' involved in structural realism are extracted from the existing mathematical structure of a given theory, but since different theories have their own idiosyncratic mathematical language, the resulting structures are highly theory-dependent and it can be difficult to make meaningful comparisons across different theories. 
 
In particular, Ladyman and Ross \cite{Ladyman2007-LADETM} have argued that structural realism should be regarded as simply constructive empiricism combined with a belief in objective modality, i.e. it differs from constructive empiricism only in a commitment to something objective that makes the equations empirically adequate and grounds the expectation that they will continue to work in the future. But many extant structural realist approaches seem to diverge from that vision by extracting structure directly from the mathematical models of an existing theory and then simply disregarding the theoretical entities which are usually supposed to instantiate that structure, and one might worry that this only superficially circumvents the epistemic difficulties of scientific realism. The approach we have proposed here offers an alternative route to cashing out  the vision of Ladyman and Ross - for as we have already noted, the operational theories approach has in common with constructive empiricism the idea that we should refrain from making commitments to theoretical entities, and thus by insisting on a completely operational construction we are performing a much more thorough rejection of theoretical entities than other structural approaches which simply discard objects from an existing object-oriented model. But on the other hand, this approach also goes beyond constructive empiricism in making an explicit commitment to a unique objective modal structure which can't simply be read off the Humean mosaic - the claim that there is some underlying reason \emph{why} quantum mechanics empirically adequate is distinctly outside the scope of constructive empiricism. Thus this approach allows us to make a metaphysical commitment which goes beyond the empirical without explicitly or implicitly invoking theoretical entities of any kind, thus fulfilling the mandate captured in the slogan  `just constructive empiricism plus modal structure.'  We therefore  circumvent the common objection that structural realism collapses either into standard scientific realism on the one hand, or into constructive empiricism on the other\cite{pittphilsci14116,newman1928mr,Psillos1995-PSIISR-2}: we avoid the collapse into standard scientific realism by insisting on a completely operational construction, and we avoid the collapse into constructive empiricism by insisting that even when we have several structures achieving equal empirical adequacy, there is some objective fact of the matter about which of these structures gets closest to the actual objective structure of reality. 
 
Another long-standing objection to structural realism involves the suggestion that it makes no sense to believe in structures without also believing in objects that instantiate those structures. And this may indeed seem somewhat puzzling  if our route to structural realism involves taking some standard object-oriented theory and simply removing all the objects, because it's unclear what  structures of that sort represent without the objects that are typically supposed to instantiate them\cite{Chakravartty1998-CHAS-20,Psillos2001-PSIISR}.  But the form of structural realism we have presented here avoids this problem, because it does not arrive at structure by simply removing the objects  from a standard object-oriented model. The structures in question are modal structures defined in terms of sets of Humean mosaics,  which are understood (roughly speaking) as distributions of categorical properties across spacetime. No theoretical entities were invoked at any stage during the construction of these structures and therefore there is no need to go through a step of removing the theoretical entities or to puzzle over what is left behind once the theoretical entities are eliminated. 
 
Finally,  it is common to object to structural realist approaches on the grounds that a lack of clarity regarding the nature of the structures in question  leads to some unsatisfying vagueness - for example Esfeld argues that OSR is only a `partial realism' because it provides `only a general scheme for an ontology of the physical domain,' and hence it is too vague to say anything interesting about, for example, how the EPR correlations come about\cite{articleEsfeld}. Identifying  axiomatisations with potential structures offers one promising way to answer such objections:  the structure in question is modal structure and the operational theory framework provides us with a clear and precise way of capturing features of objective modal structure, with different axiomatisations representing different proposals for the modal structure which might be associated with a given theory. In particular, in answer to Esfeld's concerns about the EPR correlations we can provide a number of different axiomatisations of quantum mechanics, each offering a slightly different story about the reasons why the EPR correlations take the form that they do.

We reinforce that the structural realist approach we have proposed here is not incompatible with the belief that there exists some underlying mechanistic explanation of quantum mechanical behaviour, e.g. an interpretation of some kind, or a grand unified theory, or quantum gravity: this would simply have the consequence that our approach should be regarded as a form of \emph{epistemic} structural realism, where we maintain that the structures in question are instantiated by some substratum but we don't have epistemic access to that substratum.  In this case the constraints induced by the laws of nature are to be understood as applying most directly to the elements of the underlying substratum, and then these constraints will naturally have consequences for operational features of objects composed out of these elements, so we will still be able to obtain their support on the set of OT-mosaics, i.e. the set of constraints $\{ C' \}$. On the other hand, one could certainly also pursue a more radical ontic structural realist approach which insists that constraints like `no-signalling' are themselves genuinely fundamental and there is no mechanistic substratum which gives rise to these constraints. On this view, the modal structure itself is the complete and final explanation for quantum behaviour (modulo the need to provide some answer to the measurement problem, as discussed in section \ref{counterfactual}; it would certainly be interesting to consider the extent to which a structural approach might have the resources to address that question, but that is somewhat outside the scope of this paper).

 \subsection{The Pessimistic Meta-Induction \label{pm}}

It has been suggested that research in the operational theories framework can be regarded as a way of circumventing the pessimistic meta-induction - for example, Mittelstaedt presents a systematic account of theory change as a process of revisions of axiomatisations\cite{mittelstaedt2010rational}. Now, if the intention here is to argue that because certain features of quantum mechanics can be derived from `fundamental' principles they are guaranteed to survive theory change, this argument seems naive: as noted in section \ref{counterfactual}, apparently `fundamental' principles can be and often are overthrown in scientific revolutions. And in any case, this argument cannot even get off the ground without some specification of how exactly we should go about selecting the `simple, fundamental' principles which we expect to survive theory change, since there are an infinite number of different ways of `axiomatising' quantum mechanics. 
 
But the approach we have set out here provides one way of cashing out this argument: axiomatisations are \emph{hypotheses} about the way in which quantum mechanics arises from the underlying modal structure of reality, and therefore they are intended to help us understand how it could be that quantum mechanics has such great empirical success even though the pessimistic meta-induction gives us reason to suspect that the theory itself is probably not strictly correct. Indeed, we could easily imagine future philosophers of science looking back and saying something like `Well, although quantum mechanics was not strictly correct, we now understand that it works in certain domains because it follows from the principle of the existence of a unit of information etc, and those principles also hold in the relevant domains according to the successor theory which we now believe.' This is similar to the sort of analysis which today's philosophers of science apply as they seek to understand why the theories of the past were successful despite their flaws (e.g. see ref \cite{doi:10.1086/670297}).
 
 Note that it is important here that we take a \emph{non}-selective approach, because it is relatively unlikely that \emph{all} of the  axioms that have been used in various places to derive quantum mechanics will be preserved across theory change, and this threatens to derail a selective approach to realism for axiomatisations. For example, ref \cite{pittphilsci14116} objects to modal approaches to structural realism on the grounds that modal structure  often fails to be preserved through theory change; but of course this is a pertinent objection to modal structural realism only if it is understood as a form of \emph{selective} realism where we are supposed to be able to identity  in advance which particular modal structures are responsible for the success of our theories. The non-selective modal structural realist has no problem here: they are free to acknowledge that modal structure may appear to be lost during theory changes because we don't necessarily know precisely \emph{how} the theory is embedded into the objective modal structure of reality, and therefore we may simply be wrong about the modal features we assign to our theories.  Thus if we wish to invoke the OT approach in response to the pessimistic meta-induction, it is in fact important to regard different axiomatisations as genuinely distinct possibilities, as a reflection of the epistemic humility associated with the non-selective account of the success of scientific theories. We have good reason to explore a variety of different ways in which the theory could be embedded in the modal structure of reality in order to have a better chance of coming upon the true explanation of the success of the theory, though we will only know for sure what that explanation is once we arrive at a successor theory and are able to look back and see which of the axioms proposed in various different axiomatisations of quantum mechanics have in fact survived.

 \subsection{Principle Theories}
 
 An important taxonomy of physical theories was set out by Einstein, who distinguished between what he referred to as  \emph{principle theories} and \emph{constructive theories}\cite{Einsteinwhatis}. A principle theory involves using a set of well-confirmed empirical generalizations to make inferences about some phenomena, whereas a constructive theory involves setting out a specific model for the reality underlying the phenomena. For example, thermodynamics is usually regarded as a principle theory,  whereas statistical mechanics is usually regarded as a constructive theory. Einstein saw principle theories as inferior to constructive ones, and believed that the purpose of principle theories is simply to  make progress in periods where we have not yet been able to come up with a constructive theory. Indeed, he described his own theory of special relativity in this way.\cite{brown1999origin} 
 
 Several authors have suggested that operational axiomatisations of quantum mechanics can be regarded as principle theories\cite{Koberinski,bub1999quantum,Masanes_2011}. However, ref \cite{Grinbaum_2017} observes that these principle theories aren't doing exactly what Einstein intended, for Einstein saw his principles as governing the behaviour of physical entities, whereas the OT approach very deliberately abstracts away from any specific physical entities. Indeed, in a sense standard quantum mechanics already gives a detailed description of microscopic phenomena, so by Einstein's standards we  seem to be going in the wrong direction by moving from a constructive theory to a principle one.
 
 Moreoever, the original concept of a principle theory doesn't really  seem to accommodate the idea that some empirical generalizations might be better or more fundamental than others. Einstein observed, in a letter to Solovine\cite{einstein2011letters} that \emph{`there is no logical route leading from (immediate experience) to (axioms),'} so he was certainly aware that there may exist alternative sets of principles which would give rise to the same empirical predictions, but in this letter there is no suggestion of how we ought to evaluate different axiomatisations against one another. And after all if principle theories are just stop-gap theories designed to allow us to predict some things without having a constructive theory for the underlying reality, there would be little point in coming up with alternative principles which lead to the same predictions, since the axioms are to be regarded as simply a pragmatic toolbox for making predictions.  
 
 But the recent enthusiasm for new axiomatisations of quantum mechanics seems to be doing something very different. Grinbaum suggests that what we have here is a movement toward taking principle theories  more seriously - i.e. regarding them as complete theories in themselves rather than stepping stones toward some final theory\cite{10.1086/525620}. And if that's the case, Einstein's original rationalization for principle theories is no longer adequate, so we seem to be in need of a new account of the nature and purpose of principle theories. Regarding them as an implementation of structural realism can provide such an account. In this picture, the original constructive theory represents standard object-oriented realism, and the principle theory represents the structural realist version - it makes the same predictions and is still intended as a description of objective reality, but it has less ontological baggage and focuses on modal structure rather than objects and systems.

 \section{Outlook \label{future}}
 
According to the interpretation we have suggested here, the operational theories framework offers a precise way of studying modal structure without making too many assumptions about ontology, which makes it an ideal setting in which to pursue and formalise structural realist intuitions. This suggests that the approach taken within the OT framework could be more generally applicable, and therefore  we would ideally like to employ the OT framework in domains beyond quantum mechanics. 
 
The good news is that there already exist operational formulations of General Relativity. For example,  Ehlers, Pirani, Schild have set out an axiomatic operational approach to space-time where light rays and test particles under free fall are taken as primitives\cite{EPSrpub};  De Felice and Binwe have studied how to model measurements in GR\cite{de2010classical}; and Hardy has given  an operational-tensor formulation of GR\cite{hardy2016operational}. It would certainly be an interesting project to see if the structural realist interpretation of OTs that we have given here can be extended to these operational approaches to GR; the result would potentially be a valuable contribution to the existing literature on structural realism about spacetime\cite{pittphilsci2778,pittphilsci2873}.  Hardy has also set out an operational version of quantum field theory\cite{hardy2016operational} and  made some suggestions regarding approaches to an operational version of quantum gravity.

That said, it is not entirely straightforward to apply the framework we have proposed here to these other operational theories, because  `Humean mosaics' are typically understood in terms of distributions of categorical properties across spacetime, so we will be unable to use mosaics of that kind in the context of GR where we are not supposed to assume the existence of a fixed static spacetime on which properties are distributed. Similar problems arise for quantum gravity, because several important approaches to quantum gravity have the consequence that spacetime is not fundamental but rather emergent from some underlying substratum. However, we can deal with these possibilities by appeal to a construction similar to the one we have already used for OT-mosaics. First, we replace the space of Humean mosaics with the space of  generalised mosaics, where a generalised mosaic is still defined as the set of all local matters of particular fact within a possible world, but we now allow that these local matters of particular fact are not simply facts about distributions of categorical properties over spacetime, but instead may include facts about the fundamental substratum out of which spacetime itself emerges - for example,  categorical properties across a (non-spatiotemporal) operational causal structure, as in the process matrix formalism\cite{Oreshkov2}. Then we can define a subset of the  generalised mosaics that we will refer to as the quasi-Humean mosaics - that is, the set of generalised mosaics in which there exists something which is at least approximately a spacetime and thus can be associated with something which is at least approximately like a Humean mosaic. For example, if the fundamental stuff turns out to be causal sets, the quasi-Humean mosaics will be those whose causal sets reproduce a four-dimension spacetime in the appropriate limit - the causal set research programme has devoted considerable effort to developing `dynamics' that select the approprate set of causal sets\cite{article}. In this picture, the constraints induced by the laws of nature are defined as sets of generalised mosaics, but given that the actual mosaic is clearly a quasi-Humean mosaic, we can infer that each of the laws of nature must have some support on the set of quasi-Humean mosaics, and thus we obtain a set of quasi-Humean constraints given by the support of the constraints induced by the laws of nature on the set of all quasi-Humean mosaics. Then if we want to understand how the axiomatisations of QM studied in this paper can arise from this more general context, the constraints consisting of sets of Humean mosaics that we employed in our original analysis can then be identified with the quasi-Humean constraints, i.e. they are equal to the support of the constraints induced by the laws of nature on the set of all quasi-Humean mosaics. The rest of the construction then proceeds as before: the quasi-Humean constraints in turn induce a set of constraints with support only on the set of OT-mosaics and these constraints can be used to extensionally define the axioms in an axiomatisation of quantum mechanics. 

Note that there is a possible objection to the general applicability of the OT formalism arising from the fact  that it takes as primitive the concepts of `preparation,' `transformation' and `measurement,' and also makes certain assumptions about composability, space, time and so on.  This means that the framework necessarily builds in a distinction between agent and world, so it's not straightforward to model agents and their actions within that framework.   For this reason, it seems likely that the OT framework must be thought of an interim solution, and we should ultimately be working towards a similar `operational' framework which does not depend on a distinction between agent and world. For example, Oreskhov and Cerf\cite{Oreshkov2} have set out the process matrix formalism, which allows us to generalize the framework of operational theories in a way that does not depend on a predefined time or causal structure, thus giving us the mathematical resources to deal with theories that might contain indefinite causal order, causal loops or other structures that don't fit into our familiar notions of time and causality.  It seems likely that as time goes on research on OTs will increasingly be replaced with research in the process matrix framework or some other such generalisation. That change in emphasis will most likely provide additional support for the interpretation we have presented here: as operational approaches become less focused on the specific way in which we ourselves experience the world, it will likely become important to appeal to the existence of \emph{objective}, mind-independent modal structure in order to make sense of this sort of perspective-free approach.

 \section{Conclusion} 
 
Derivations of quantum mechanics from a set of simple, empirically well-confirmed statements have some intuitive plausibility as explanations of why quantum mechanics is the way it is.  But there are many different empirically well-confirmed statements from which we can derive quantum mechanics, many of which will appear simple from some point of view, and therefore in order to make sense of the idea that certain axiomatisations confer insight while others do not, it is necessary to take some view on the epistemic and/or ontic status of the axioms, i.e. to provide an interpretation of them. 
 
In this article we have set out an approach which aims to accommodate this research programme in a scientific realist worldview. First, we have suggested that the relation `every theory which exhibits feature $A$ must also exhibit feature $B$' is most naturally understood as expressing metaphyiscal necessitation, and therefore an axiomatisation of quantum mechanics consists of setting out a collection of axioms which are together sufficient to metaphysically necessitate quantum mechanics. Second, we have suggested that if we adopt a realist picture which incorporates governing laws (as opposed to merely descriptive Humean laws), it follows that the laws of nature impose a set of constraints on OTs, and thus there is some `natural' axiomatisation induced by the laws of nature which encodes the objective modal structure of reality. Thus scientific realists can understand axiomatisations of quantum mechanics as hypotheses about this objective modal structure. As this interpretation involves putting forward  hypotheses about structure which go beyond the empirical data, it is a form of realism, and since it involves working directly with  modal structure without making any claims about the entities which instantiate these structures, it is a form of structural realism, although not  \emph{selective} structural realism.
 
This realist interpretation has a number of important consequences. First it clarifies the sense in which we can obtain insight from axiomatisations of quantum mechanics. Second, it explains why certain sorts of axiomatisations are preferable. And perhaps most importantly for the practising scientist, it holds out hope that research on operational theories and new axiomatisations of quantum mechanics might do more than just give us a subjective feeling of understanding - if axiomatisations are hypotheses about the constraints on the space of  OT-mosaics induced by he laws of nature, then finding the correct axiomatisation could allow us to make backwards inferences about the laws of nature, which may help guide future research programmes.  Thus this account allows us to acknowledge the scientific importance of research on operational theories; we engage in this research not merely to satisfy our intellectual curiosity but also in the hope that it will facilitate scientific progress. 

Finally, this approach also has consequences for philosophy of science more broadly. In particular, many influential approaches to structural realism suggest that we should be realists about modal relations, but it is not always clear which particular modal relations are involved or how these modal relations should be expressed\cite{doi:10.1080/02698590903006917}. The OT framework offers one possible way of answering these questions, by placing quantum mechanics within a wider space of theories in which we can do modal analysis using a modified version of the possible world semantics. It also enables us to acknowledge the profound importance of operational thinking in physics without obliging us to go down the positivist route that has come to be associated with the term `operationalism.' Thus is seems likely that the tools developed for research in the OT framework and related frameworks may be valuable to structural realists as a means of cashing out the modal structures to which they are committed, and so structural realists would do well to make use of the developments in this  increasingly active field of research. 

\section{Acknowledgements} 

This publication was made possible through the support of the ID 61466 grant from the John Templeton Foundation, as part of the “The Quantum Information Structure of Spacetime (QISS)” Project (qiss.fr). The opinions expressed in this publication are those of the author(s) and do not necessarily reflect the views of the John Templeton Foundation.

  \bibliographystyle{plainurl}
 \bibliography{newlibrary12}{}

 \end{document}